\documentclass[%
 aps,
superscriptaddress,
groupedaddress,
showpacs,
 amsmath,amssymb,amsthm,mathrsfs,
prl,
twocolumn
]{revtex4-1}

\usepackage{graphicx}  % needed for figures
\usepackage{dcolumn}   % needed for some tables
\usepackage{bm}        % for math

\graphicspath{{Pictures/}}

\begin{document}

%\preprint{APS/123-QED}

\title{Morphological Inversion of Complex Diffusion}% Force line breaks with \\

\author{V.~A.T.~Nguyen}
 \affiliation{University of Notre Dame, Department of Physics, 225 Nieuwland Science Hall, Notre Dame, IN 46556 USA}%Lines break automatically or 
 %can be forced with \\
\author{D.~C.~Vural}%
\email{dvural@nd.edu}
\affiliation{University of Notre Dame, Department of Physics, 225 Nieuwland Science Hall, Notre Dame, IN 46556 USA}%

\begin{abstract}
Epidemics, neural cascades, power failures, and many other phenomena can be described by a diffusion process on a network. To identify the causal origins of a spread, it is often necessary to identify the triggering initial node. Here we define a new morphological operator and use it to detect the origin of a diffusive front, given the final state of a complex network. Our method performs better than algorithms based on distance (closeness) and Jordan centrality. More importantly, our method is applicable regardless of the specifics of the forward model, and therefore can be applied to a wide range of systems such as identifying the patient zero in an epidemic, pinpointing the neuron that triggers a cascade, identifying the original malfunction that causes a catastrophic infrastructure failure, and inferring the ancestral species from which a heterogeneous population evolves.
\end{abstract}

\maketitle

A sugar piece placed in tea will erode and eventually dissolve. Given the initial shape of the piece, it is trivial to predict its final distribution. However, the opposite problem of determining the initial state, given a final one is extremely difficult. Problems of the latter kind are referred as ill-posed inverse problems \cite{inverseBook, gaussianDeconvolution, imageDeconvolution}. 

Diffusion taking place on networks, in the forward direction, is well studied. One class of models originally used to describe epidemics is the Susceptible-Infected-Recovered (SIR) model \cite{epidemicModels, KermackSIR}. Variations include SI, SIS, SIRS, etc. Others include more realistic delay conditions, such as an incubation period for the infection \cite{incubation}. Similar models are used to describe neural cascades \cite{neural}, traffic jams  \cite{trafficDiffusion} and infrastructure failures \cite{powerFailure}. 

Accordingly, a successful method of inverting diffusion on complex networks can help identify patient zero in an epidemic outbreak, pinpoint neurons that trigger a cognitive cascades, remedy the parts of the road network that initiate congestion, and determine malfunctions that lead to cascading failures. In the weak selection limit, evolution can be thought as diffusion on a genotype network \cite{neutral1,neutral2}, so diffusion inversion may be used to identify ancestral species.

Here we address the problem of identifying the origin of a diffusive process taking place on a complex network, given the its final state. We refer to the influenced nodes as the candidate set $C$. Any member of $C$ may be the node from which the diffusion originated. We refer to this node as the seed, $s$, and to the forward model as $M$.
%Intuitively speaking, the difficulty of inverting any diffusive processes stem from a combination of factors: The first is the interaction of the spreading agent with the \emph{boundaries}; e.g. consider a diffusive process taking place within a cylindrical domain, with finite radius but infinite height. Shortly after the spread reaches the rim, estimating the horizontal coordinate of the diffusion origin becomes far harder than its vertical coordinate. The second difficulty stems from \emph{self interactions}, For example for a diffusive process taking place on the surface (not volume) of a cylinder, once the spread wraps around, estimating the azimuthal ($\phi$) coordinate of the diffusion origin will be far easier than estimating its vertical ($z$) one. The third difficulty, which amplifies the severity of the first two- is \emph{stochasticity}. For example the origin of a random walk is harder to estimate than a deterministic trajectory. 

%Inverting diffusion on a network embodies all difficulties listed to the worst extent: In a typical random network, the diffusion front frequently runs into ``boundaries'' (dead ends), where the spread gets saturated. Secondly, random networks contain a large number of cycles frequently leading to self interactions. Lastly, realistic network diffusion models always involves some degree of randomness. As a result, past attempts of inversion has been very model specific. \cite{bayesBelief, siri, temporalRobust, rumorSITreeDong, rumorCentrality, rumorreeShah, rumorTimeVarying, virusDistGraphs, noisy, DMP}.

Presently, there are two approaches to identify $s$. The first uses probability marginals from Bayesian methods \cite{bayesBelief, siri, temporalRobust, rumorSITreeDong, rumorCentrality, rumorSITreeShah, rumorTimeVarying, virusDistGraphs, noisy, DMP}. In some cases, it is possible to sample the state space using Monte Carlo simulations \cite{temporalRobust}. However, this is only feasible for small networks. Message-passing algorithms can approximate the marginals efficiently \cite{bayesBelief, virusDistGraphs, noisy, DMP}, however these algorithms are  model specific: for every $M$, one must invent new approximations, heuristic assumptions and analytic calculations.

\begin{figure}
\centering
\includegraphics[width=\linewidth]{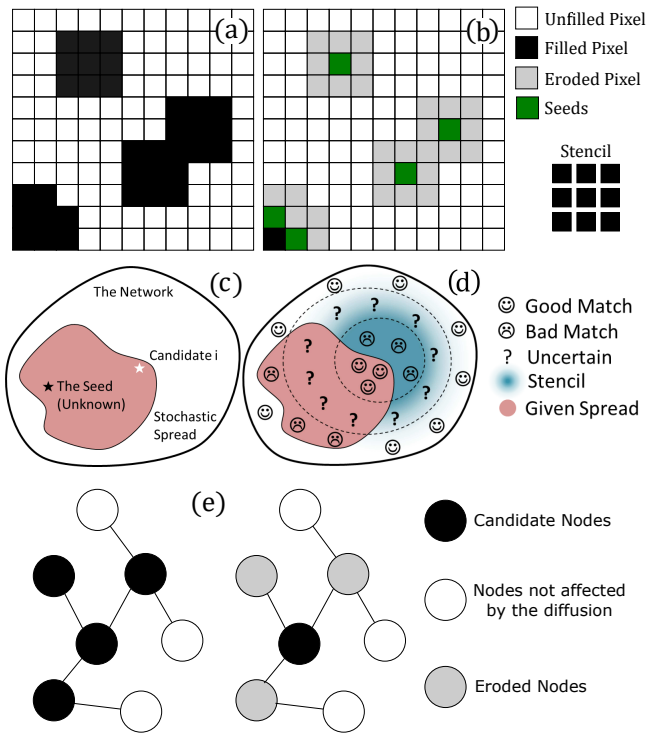}
\caption{{\bf Example of classical erosion (a,b), our generalization (c,d), erosion on a network (e)}. (a): The structuring element (``stencil'') is placed on individual pixels whose neighborhoods are checked for a match/mismatch. (b): Pixels are eroded if there is a mismatch with the stencil. In the end, shapes lose their outer layers. (c): the network and candidate set (solid fill) is given while the seed (dark star) is unknown. (d): the erosion stencil (gradient fill) for candidate i (light star) is applied over the network. Smiley faces show the locations where the stencil matches the given network state, question marks show locations of high variability, and sad faces show regions where the stencil does not match the candidate set. Note that the planar representation for the stencils does not mean that our stencils are limited to planar graphs. All nodes are part of the stencil for any given node for a general diffusion dynamic. (e): An example of deterministic diffusion of a single seed for \emph{one} discrete time step. The necessary stencil is one in which all neighbors are affected by the diffusion but no one else because there is not enough time to reach anything outside of direct neighbors. For this reason, the node in the top left was also eroded.}
\label{fig:scheme}
\end{figure}

In contrast, the second class of methods works independent of the forward model \cite{rumorSITreeDong, rumorCentrality, rumorSITreeShah, virusDistGraphs}. These presuppose that $s$ should be approximately equidistant to all other nodes in $C$, and therefore, nodes with high ``centrality'' values should have a higher likelihood of being $s$. This assumption breaks down if the spread reaches ``boundaries'', or if the spread self-interacts (i.e. if the network contains many loops rather than being a tree or a dynamic like SIS). 

% \begin{figure*}
% \centering
% \includegraphics[width=\linewidth]{protSIde_2.pdf}
% \caption{Error distance $\delta$ for the protein interaction network ($n=3744$ and $m=7749$) using our method/distance centrality/Jordan centrality (left to right). $\delta$ is the distance between the top candidate and the true origin.}
% \label{fig:DPROT}
% \end{figure*}

Here we present a method that can determine the origin of a diffusive process taking place on a complex network, regardless of what the diffusion model is, without the drawbacks of centrality based methods. We take as inputs the network structure, the candidate set $C$, and the forward model $M$. In return, we output a list of nodes, ordered according to the likelihood of being the seed $s$. We emphasize that our method has no free parameters and is applicable to any $M$, including both deterministic and stochastic forward models. 

To evaluate our success, we performed simulations in the forward direction using 4 types of forward models on 3 different graph topologies. We then inverted the final state, and determined how often our guess is the true seed. We also measured the error distance, i.e the distance of our guess from the true seed.

The forward models we explored are Susceptible-Infected (SI) epidemic model with uniform propagation probability between neighboring nodes; an Information Cascade (IC) model which propagates the diffusion like the SI model but with an additional cascade effect based on the fraction of infected neighbors \cite{WattsCascade}; a Collective Behavior (CB) model based on the notion that social behavior is determined by threshold for when the benefit of an action is greater than its cost \cite{GranoThreshold}; a heterogeneous SI model where the propagation probability has a directional bias (DB) based on spatial positions of the nodes. 

The network topologies on which we evaluate our model consist of a real power grid (GRID) network of the western states of the USA \cite{powerGrid}, a real protein (PROT) interaction network of C. elegans \cite{biogrid}, and a synthetic scale free (SCLF) network based on the power grid network.

%The spread time was selected so that the number of affected nodes were as large as possible without the diffusion process filling the entire network.
The spread time was selected such that none of the networks tested was fully saturated by the spread. This allows the final state to retain some unique characteristics that can be used to identify the seed. We did not explore cases for low propagation probability and large spreads in order to maintain a consistent total simulated time, $T=5$, for all models. 

{\bf Generalized Morphological Operators.} The principle behind our method can be best described by the language of mathematical morphology pioneered by Minkowski, Matheron, Serra and others \cite{morphology, mathmorph, minkowskiOperators}. A morphological operator modifies every point in a set (e.g. an image) according to the spatial arrangement of neighboring points. A stencil, called the ``structuring element'', with a predefined shape is placed on individual points, and if the surroundings of the point match (or not match) the shape of the stencil, then the point is modified. One particular operator, ``erosion'', is important for our purpose. Erosion, deletes all points whose surroundings mismatch the structuring element. Since a mismatch would typically happen near the boundaries of a shape, the erosion filter ends up rounding up and thinning down all shapes. This is the qualitative behavior we need in order get rid of the peripheral nodes of $C$ and reach its core.

To suit our specific purpose, we define a new morphological operator analogous to erosion, but with three important differences (Figure \ref{fig:scheme}). First, our structuring element is not fixed, but changes according to where it is placed on the network. Furthermore, our structuring element does not have sharp edges, but is fuzzy. To be precise, we take the structuring element, when placed on $i$, to be $P(j|i)$, which we compute numerically.

Secondly, the comparison of the structuring element and the surrounding nodes of $i$ is not binary, but weighted. This is because mismatches of deterministic events (e.g. $P(i|j)\sim0 \mbox{ or } 1$) matter more than random events ($P(i|j)\sim0.5$). To be precise we weight every node mismatch with a factor inversely proportional to the binary entropy $H_b(P(j|i))$.

Thirdly, the final effect of processing a node with a structuring element is not simply deleting or keeping. Instead, this too is fuzzy. In the end, upon applying our morphological operator to the network once, we expect the least eroded node to be the seed.
%Here we define a new kind of morphological operator to suit our purpose. Our morphological operator has two important features that are fundamentally different than conventional ones. {\bf(1)} We define an adaptive structuring element. This means that the form of the structuring element depends on where it is placed on the network. Specifically, we take the structuring element to be $P(j|i)$ when placed on $i$. We obtain $P(j|i)$ by running the forward model multiple times, setting $i$ as the seed. {\bf(2)} The mismatch criteria is not binary, but fuzzy. Qualitatively speaking, the mismatch when $P(i|j)$ is close to 0 or 1 matters far more than a mismatch when $P(i|j)$ is close to $0.5$. Specifically, we weight every match / mismatch with a factor proportional to the self information (surprisal) $I_{ij}=\log1/P(i|j)$ of the event. 

Our algorithm generates an ordered list of candidate seeds based on how much they are eroded. The best case scenario is when the true origin is located at the top of this list. Fig. \ref{fig:scheme} schematically shows an evaluation of the match between an erosion stencil and the given network state along with an example of classical image erosion.

{\bf Method.} The structuring elements are generated by directly sampling the states of the network model. For each forward model (defined explicitly in the next section), we applied the selected diffusion dynamics 500 times for every node in the network and calculated the structuring element, $P(i|j)$, based on the normalized frequency of finding node $i$ affected by the diffusion starting from node $j$. %Increasing the sample size from 500 up to $10^4$ for selected nodes only gave marginally better results.

We define a convergence condition for our stencils based on the average absolute error of the probabilities after moving to a higher sample size. For a sample size $n$, we require that the average error, $\delta P$, be less than $1\%$ after moving to a sample size of $5n$:  \[\delta P = \max_{\lambda \in (0.1,0.5,0.95)} \sum_{i,j}|P^{5n}_\lambda (i|j)-P^{n}_\lambda (i|j)| \leq 0.01.\] Naturally the error will depend upon the diffusion parameters and the network topology. However, we found that it is reasonable to just sample the errors using the largest network for a subset of the diffusion parameters. For the most part, our algorithm convergences very quickly where the final sample size for the stencils was $500$ runs for each node in the network.

The idea behind our morphological filter is to determine which stencil has the best match with the given diffusion state. The likelihood of a node being the origin node is proportional to the similarity between its stencil and the diffusion state. However, there are many ways to measure the similarity between a probability vector and a binary vector. We sampled the performance of different scoring metrics such as log-likelihood (under independent 3-body correlations), information surprisal, and Picard distance. The results shown in this paper are based on the an information theoretical metric which we found to work best. For each node $i$ inside the candidate set $C$, we apply an erosion score:
\[S_{i}=\sum_{j} \frac{1-P(j|i)}{H_b(P(j|i))}, \]
where the weight $H_b(x)=x\log_2 x+(1-x) \log_2(1-x)$ is the binary entropy. In other words, $S_i$ measures the mismatch between the stencil and diffusion state weighted by the (binary) entropy, $H_b$, of the probability distribution. This weight will diminish the value of nodes with high variability ($p \approx 0.5$) in comparison to nodes with low variability ($p \approx 0$ or $p \approx 1$). If the probability that a node is affected is very high or very low, then matches (mismatches) are weighted heavily and have a large negative (positive) influence on $S_i$. On the other hand, the state of highly variable nodes are circumstantial, and thanks to the small entropic weight they do not have much influence on $S_i$. Note that the score for candidate node $i$ examines its stencil element, $P(j|i)$, at every other node $j$. In other words, the stencil for any node involves all other nodes in the network and not just the nodes in the set $C$.

Once we have $S_i$ for all $i$ we sort these in ascending order and pick the nodes with best (i.e. lowest) scores. Numerically, the entropic weight can result in a division by zero, and therefore instead of directly using $P(j|i)$ we used $P(j|i)+\epsilon$, where $\epsilon = 10^{-20}$ is small enough to not change the degree of variability and instead provide an upper bound for a highly unexpected mismatch. If the relevant forward dynamics is one where nodes can take more than two states, then the binary entropy function should be updated to be the entropy for the probability stencil of node $i$ causing node $j$ to be in state $r$:
\[ H = \sum_r P(r,j|i)\log_e P(r,j|i)\]

The error bars generated are based on the standard deviation of our results. For each network topology and diffusion model we simulated 500 realizations of the diffusion in order to test the performance of our algorithm. We separated these into 5 sets of 100 simulations by random assignment. We then calculated our performance inside each of the sets separately and used their standard deviation for our errors. We then repeated the random assignment 100 times and averaged over all of the standard deviations. This removes the dependence on how the simulations were randomly assigned as well as provide a measure of ``error'' for a trial containing 100 simulations. Additionally, some realizations will not have a candidate set larger than one when the spread probability is very small. We simply remove these cases from our calculation and therefore the actual sample size for low $\lambda$ ($\approx 250$) is lower than for $\lambda > 0.80\%$ (500).

{\bf Networks and Forward Dynamics.} To evaluate our inversion scheme, we used a protein-protein interaction network \cite{biogrid}, a power grid network \cite{powerGrid}, and a synthetic scale-free network. Our diffusion dynamics are discrete in time and are fully described by the probability that an ``infected'' node spreads to a susceptible neighbor. The SI model is defined by the probability $p_\textrm{ij}=\lambda A_{ij}I_j$ of an infection spreading from $j$ to $i$, where $A_{ij}$ is one when nodes $i$ and $j$ are connected by an edge and zero otherwise and $I_j$ is one if node j is infected and zero otherwise. Which simplifies to $j$ infecting $i$ with probability $\lambda=[0.05,0.1,...,0.95]$ only if the two nodes are adjacent and $j$ is infected.

The IC model cascades the information spread based on the state of a critical fraction of neighbors, $\nu=0.5$, via:
\begin{equation}
p_{ij}=\begin{cases}
	1, & \sum_{j} A_{ij}(I_j - \nu) \geq 0,\\
	\lambda A_{ij}I_j, & \sum_{j} A_{ij}(I_j - \nu) < 0. \\
   \end{cases}
\label{SIo}
\end{equation}
The CB model spreads the adaptation of a social behavior when the number of neighbors who have adopted the behavior reaches an absolute threshold, $\mu = 2$, via:
\begin{equation}
p_{ij}=\begin{cases}
	1, & \sum_{j}A_{ij}I_j \geq \mu,\\
	\lambda A_{ij}I_j, &  \sum_{j}A_{ij}I_j < \mu. \\
   \end{cases}
\label{SIt}
\end{equation}
The DB model uses heterogeneous diffusion probabilities $p_{ij}=B_{ij}I_j$, where $B_{ij}=A_{ij}(p_0+\delta p\cos(\vec{d_{ij}}\cdot\vec{b}))$. The DB model is generated by first randomizing the three dimensional positions for all nodes placed uniformly random inside a unitary volume and then calculating the unit displacement vector, $\vec{d_{ij}}$, between graphically ($A_{ij}=1$) adjacent nodes. We then picked a unit bias vector, $\vec{b}$, pointing towards one of the corners of the volume and generated the weighted adjacency matrix $B_{ij}$ based on a neutral transmission probability $p_0=[0.2,0.4,0.6,0.8]$ and range $\delta p=0.15$.

{\bf Distance and Jordan Centrality.} The origin of a diffusion process can prima facie be expected to be found near the ``center'' of the candidate set. Thus, we compare our results with two benchmark methods based on centrality measures (Fig.2). Centrality based methods calculate the distances between pairs of candidate nodes $(i,j)$, $D_{ij}$, inside the \emph{subgraph} generated by only their connections. This means that each diffusion will require a new calculation of the distances because the candidate nodes will generally never be the same set of nodes. The distance (closeness) centrality $\mathcal{D}_i$ of a node $i$ refers to the total distance between node $i$ and all other candidate nodes $j$.
\[\mathcal{D}_i = \sum_{j \in C} D_{ij}\]
This method assumes that the node which has the least distance to all other candidate nodes is the most likely seed. The Jordan centrality $\mathcal{J}_i$ of a node $i$ is concerned only with the largest distances between node $i$ and all other candidate nodes.
\[ \mathcal{J}_i = \max_{j \in C} D_{ij} \]
Similar to the distance centrality, this method assumes that the most likely candidate node is the least distant to all other candidate nodes.

{\bf Numerical Algorithm.}
Given a graph $G(V,E)$, forward diffusion model, and total time $T$, and realization $R$ to be inverted, we enumerate the necessary steps to use our algorithm.
\begin{enumerate}
\item Generate the candidate set $C$ based on $R$. For SI dynamics, the set $C$ contains every node which is in state $I$.
\item For every node $j \in C$, apply the diffusion dynamic for total time $T$ and repeat for a total of $M$ independent trails. Record the probability that node $i \in V$ was affected by the diffusion, $P(i|j)$.
\item Calculate the erosion score for every node $j \in C$ via: \[S_{j}=\sum_{i} \frac{1-P(i|j)}{H_b(P(j|i))}. \]
\item Assign ranks to every candidate node based on their scores. The lowest score has the first rank and is the most likely candidate based on our erosion.
\end{enumerate}
Our current scheme is based only on SI-type dynamics. For dynamics with additional states such as SIR, the candidate set must be carefully considered. In the most simple case, the candidate can be the entire graph. Additionally, the stencil will gain an additional dimension which categorizes the state $w$ of node $i$ for a diffusion starting from node $j$, $P(w,i|j)$. The erosion score must also use the entropy rather than the binary entropy.

\begin{figure*}
\includegraphics[width=\linewidth]{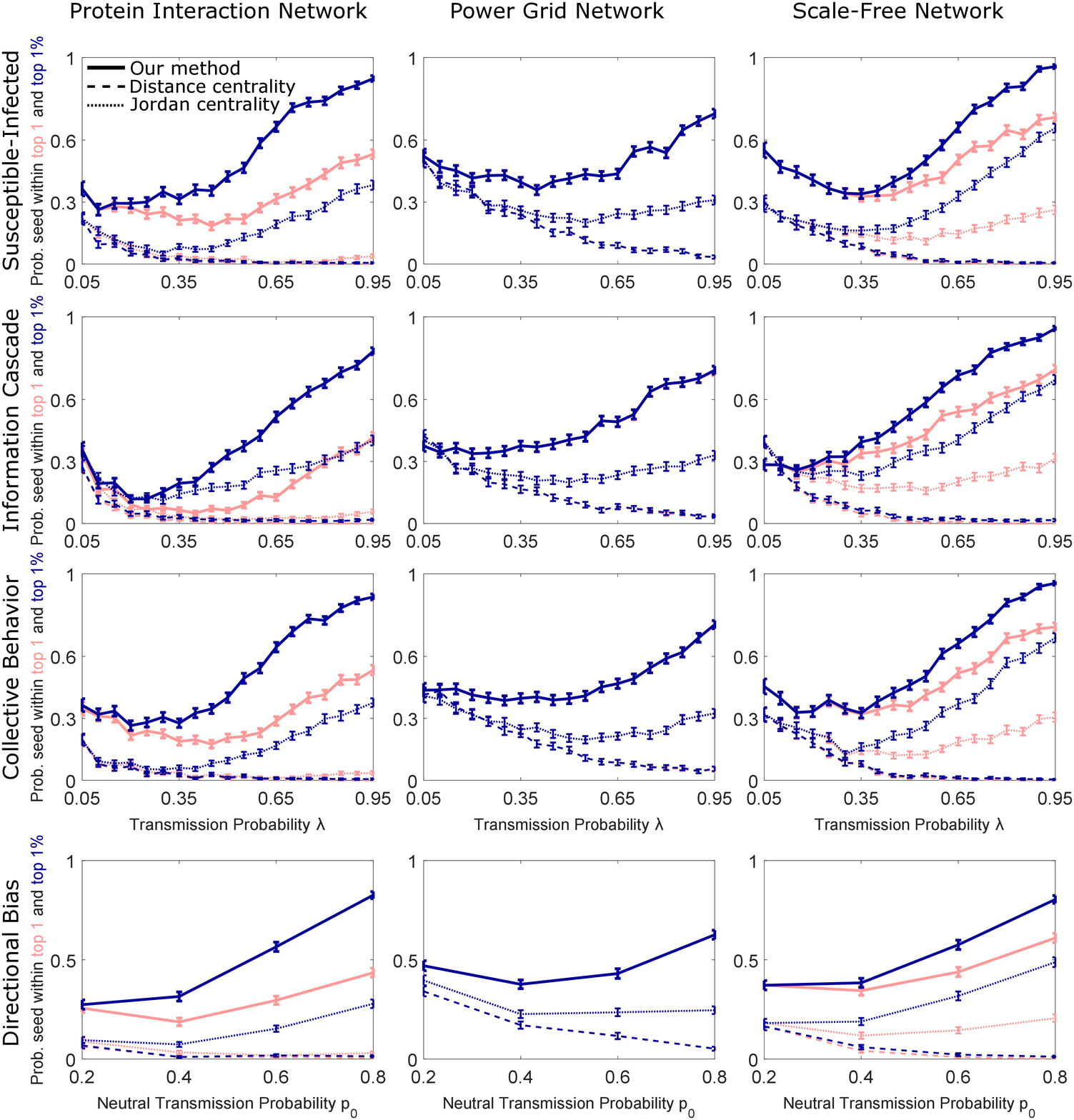}
\caption{ {\bf Performance on different networks and models as a function of the transmission probability $\boldsymbol\lambda$, for $\boldsymbol {T=5}$ (larger is better).} The protein interaction network ($n=3744, m=7749$), power grid network ($n=4941, m=6594$), and scale-free network ($n=4941, m=6601$) are shown in each column. The four dynamic models are shown in each row. Note the difference between the uniform transmission probability, $\lambda$, for isotropic diffusion and the neutral transmission probability, $p_{0}$, for anisotropic diffusion. Our performance (solid lines) is almost always better than centrality based methods (dashed and dotted lines of the same color) based on a total of 500 runs. Light-red and dark-blue curves denote whether the true seed is the top one or within top three choices returned respectively. For the power grid network, both cases overlapped because the candidate set is small. The error bars represent $\pm$ one expected standard deviation for an average of 100 runs. In general, our performance closely matches the two centrality methods but becomes noticeably better as $\lambda$ or $p_{0}$ increases.}
\label{fig:Rall}
\end{figure*}

\begin{figure*}
\includegraphics[width=\linewidth]{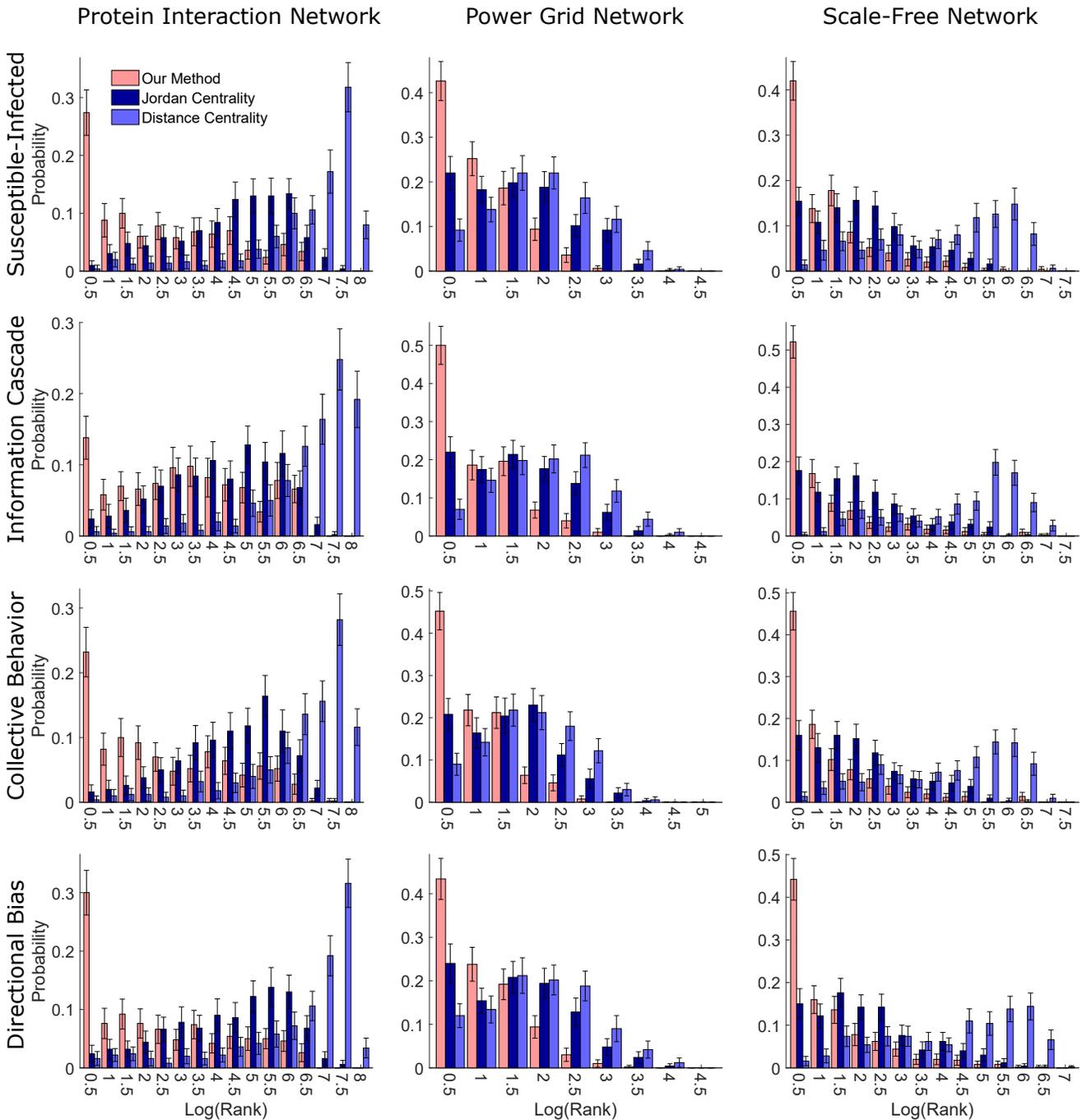}
\caption{{\bf Probability that the true seed is a high-ranked node for all models and dynamics using $\boldsymbol{\lambda=0.6}$ ($\boldsymbol{p_0=0.6}$) $\boldsymbol{T=5}$ (lower rank is better).} The protein interaction network ($n=3744, m=7749$), power grid network ($n=4941, m=6594$), and scale-free network ($n=4941, m=6601$) are shown in each column. The four dynamic models are shown in each row. The protein and scale-free networks have larger spreads than the grid network and therefore will have a larger distribution for the rank sizes. Our method (solid lines) generally outperforms the two centrality methods based on a total of 500 runs. Note that Jordan centrality provides much better performance than distance (closeness) centrality. The error bars represent $\pm$ one expected standard deviation for an average of 100 runs.}
\label{fig:rankspec2}
\end{figure*}

% \begin{figure*}
% \includegraphics[width=\linewidth]{big_normrank_2.eps}
% \caption{{\bf Normalized rank of the true seed for all models and dynamics as a function of the transmission probability $\boldsymbol\lambda$, for $\boldsymbol{T=5}$ (smaller is better).} The protein interaction network ($n=3744, m=7749$), power grid network ($n=4941, m=6594$), and scale-free network ($n=4941, m=6601$) are shown in each column. The four dynamic models are shown in each row. Note the difference between the uniform transmission probability, $\lambda$, and the neutral transmission probability, $p_{0}$. Our method (solid lines) generally outperforms the two centrality methods based on a total of 500 runs. Note that Jordan centrality provides much better performance than distance (closeness) centrality. The error bars represent $\pm$ one expected standard deviation for an average of 100 runs.}
% \label{fig:normranks}
% \end{figure*}

{\bf Results.}
%We plot the distance between the true origin and the best guess of our algorithm and compare these results to other non-model-specific methods. Overall, our algorithm performs much better than distance and Jordan centrality on the same graph type and size. Fig. \ref{fig:DPROT} shows, in a representative model and network topology, that our first guess is almost always within two hops away from the true seed. This is the case for all other models and topologies (cf. \emph{Supp. Mat}).
Many authors use distance error as a metric of success \cite{temporalRobust,virusDistGraphs,rumorSITreeShah,rumorTimeVarying}, however the usefulness of this metric is ambiguous. Although a two-hop range constitutes a small fraction of the network ($0.2\%, 1\%, 5\%$ for GRID, SCLF, and PROT respectively), such small percentages still correspond to a significant absolute number of nodes (10, 50, and 210 nodes for GRID, SCLF, and PROT respectively). We provide the results for the distance error of our algorithm in the \emph{Appendix}, and here, only focus on the probability of finding the true seed and the rank distribution of the true seed. All simulated runs were done on the Notre Dame Center for Research Computing's High Performance Computing clusters.

Figure \ref{fig:Rall} shows how often the true seed is our top guess and how often it is within our top three guesses. The bold solid lines show the performance of our algorithm, while the dashed and dotted lines show the performance of distance centrality and Jordan centrality based methods. Our success rates are far above the dashed and dotted curves of the same color, with the only exception of the low $\lambda$ regime of the IC model.

Figure \ref{fig:rankspec2} show the ranking spectrum for the true seed using the three methods for $\lambda=0.2$ ($p_0=0.2$) and $T=5$. The protein/scale-free network has a heavy tail in comparison to the grid network because the spread can quickly reach many more nodes within $T=5$ on the protein/scale-free network. 

On average the radius of $C$ is $\langle r\rangle=T\lambda$ nodes. As $\lambda\to0$, there are few nodes to pick from, and thus centrality based methods, including just randomly selecting a node from $C$, gives similarly high success rates as our method. As $\lambda$ is increased however, the difference between our method and others increase significantly. The difference in success is maximal when the number of affected nodes become maximum, at $\lambda\to1$. Across all forward models, we have least success when the spread probability $\lambda\sim0.5$, the regime with the highest number of possible states due to the high variability in the stochastic process. This variability causes the calculated average stencil to very rarely match a given diffusion state.

%For comparison purposes we have also plotted the success rate of randomly picking one node from $C$, which is one over diffusion size. If propagation probability and/or spread time is small the average total spread size is likewise small, randomly picking one of the candidates gives an acceptable result. 

%Our performance decreases as a function of time which is expected for inverse diffusion problems [cite fig:performVSTime] \cite{virusDistGraphs}. As time increases the number of initial states that can generate the final state also increases and therefore the identification of the correct initial state becomes more difficult with time even with perfect inversion.

%We can also view this relationship as the decay of information in the system over time. Given a long enough time-scale, the system becomes fully saturated by the diffusion and the stencils of all nodes converge to the same stencil. The only information remaining in a fully diffused graph is the time. Where the most graphically ``central" nodes can achieve a complete diffusion for low time-scales while the most graphically ``distant" nodes will only achieve a complete diffusion at larger time-scales. Hence, the value of the final time suggests the scope of the centrality of the origin node. However, this information becomes increasing diluted as the final time increases because eventually any node will fully saturate the network.

{\bf Discussion.} In general, stochastic dynamics on networks will be defined in terms of local properties rather than global ones. To this extent we explored two main variations to a local property, i.e. models in which fractional versus absolute number of affected neighbors determining the spread probability. We have also shown that our algorithm performs well for non-uniform and directionally-biased diffusion. Hence, we have explored, nearly to full extent, the inversion of two state diffusion processes on complex networks. 

Our algorithm is a general method for determining the source of diffusion dynamics on complex networks. While the generation of morphological stencils requires knowing the dynamic law as well as the states of all nodes at some final time, our approach works for all models. In contrast, other inversion schemes are model-specific \cite{bayesBelief,DMP,siri,rumorSITreeShah}. Even though our inversion scheme works for any model, it is not model invariant like the centrality based methods which uses only topological properties in the graph. The core part of our algorithm relies on an ``erosion'' of the diffusion surface by the diffusion stencils for each point in the spread. We need to know the specifics of the forward model in order to generate these diffusion stencils.

The performance of our algorithm may be improved by de-generalizing the scoring function to accommodate particularities of a forward model. We also note that there are many aspects of the problem we have not yet considered, such as cases of incomplete or noisy information, dynamics of multi-state diffusion, and even multi-source diffusion \cite{bayesBelief,noisy,siri,DMP,temporalRobust,rumorTimeVarying}.

Such generalizations should be within reach: in the case of multi-source, one could generate a scoring metric which is non-symmetric against the diffusion state. In other words, the scoring could prioritize matching the diffusion state rather than the base state. In the case of multi-state diffusion, one could introduce a transition matrix for the states where the elements of this matrix represents how easy/difficult it is to transition from one state to another. This matrix must be embedded into the scoring metric such that mismatches in state are weighted by the elements of this matrix. Additionally, some methods use a reduction scheme which considers more complex dynamic as two ``compartments.'' For example, \cite{temporalRobust} uses a SIR dynamic but then groups the status of I and R into the single compartment for their Jaccard similarity measure which uses a binary status.

%Even though our inversion scheme works for any model, it is not model invariant like the centrality based methods which uses only topological properties in the graph. The core part of our algorithm relies on an ``erosion'' of the diffusion surface by the diffusion stencils for each point in the spread. While any sort of propagation dynamics is macroscopically a diffusion, we have used the microscopic rules in order to generate a information dilute two body correlation which gives the big picture of the diffusion. (***sentence unclear --and probably not grammatical. Once fixed, we should try to feed the contents of this paragraph to the first and last paragraphs of this section, which already says things along these lines)

We now compare and contrast our method with another that is most similar in spirit to ours. Antulov-Fantulin et al. \cite{temporalRobust} uses a Jaccard similarity function to characterize the similarity between simulated spreads versus a given spread. This is then used to estimate the probability of a spread given a source via a Gaussian weighting and therefore we will refer to this method as the Jaccard-Gaussian algorithm. This method compares two network states, i.e. two binary vectors (nodes are either infected or susceptible) as obtained by Monte Carlo simulations versus a given final state. %This requires that their setup procedure either save all the data in order to compare to a given spread or know the given spread ahead of time. Our approach does not require knowledge of the final state before hand and therefore eliminates the need for considering the unique characteristics of each possible diffusion state. In other words,
To compare two ``micro-states'' of networks \cite{temporalRobust} must generate, store and compare all (or at least, most) possible realizations of a spread from every single candidate node. In contrast, we work with a single probability distribution defined over the network. Since the space of all possible $N$-node states (which \cite{temporalRobust} samples) is astronomically larger than the space of \emph{single} node states (which we sample) we can leverage this gain in computational cost to sample our space more accurately. %Accordingly, we do not compare two binary vectors. Instead, we compare a given final state (binary) to a probability distribution (i.e. non binary).
Our approach has another advantage: a forward model uniquely determines a stencil, and once we have our stencils for a certain model, we can use it for multiple $C$ sets, say, for different realizations of the same disease. \citep{temporalRobust} on the other hand, must realistically sample an combinatorially larger space for every realization of the spread. Another difference with \citep{temporalRobust} is in our scoring function, specifically, we use an entropic weight when comparing a single realization to a probability distribution. This allows us to decide which nodes to take more seriously than others.

\begin{figure}	
\includegraphics[width=\linewidth]{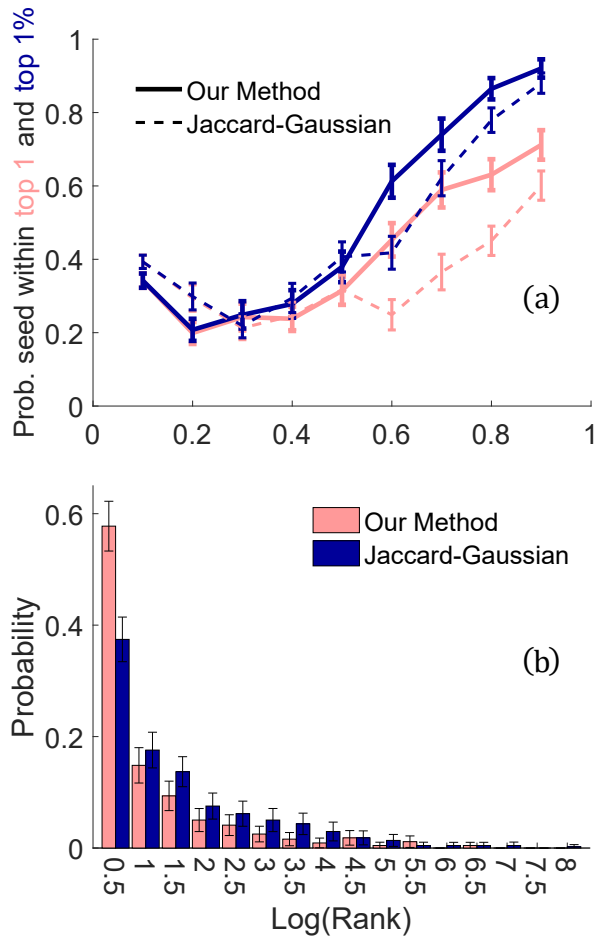}
\caption{{\bf Our method versus Jaccard-Gaussian method developed by Antulov-Fantulin et al. in \cite{temporalRobust} on the scale-free graph ($\boldsymbol{n = 4941}$, $\boldsymbol{m = 6601}$). (a): Probability of finding the true seed within top 1 (light-red) and top 1\% (dark-blue) candidate nodes.} Overall, the Jaccard-Gaussian method has better performance when $\lambda < 0.5$ after which our method is better. In all cases, both methods were limited to using the same set of 134 simulated runs (stencil) where each data point was generated from independent test sets of 500 runs. However, only runs which satisfies the Jaccard-Gaussian convergence condition (Eq. \ref{eq:JGconv}) were used in these plots. {\bf (b): How often a highly ranked node is the true seed ($\lambda = 0.7$, $T=5$).} Both methods used the same set (stencil) of 667 simulated runs for their algorithms. The spectrum is generated from testing the performance of each method for 500 runs. Additionally, the error bars were generated by randomly selecting 100 runs and calculating the deviation from average across all runs. Therefore, the error bars represent the expected standard deviation for a sample size of 100 runs.}
\label{fig:Ent_JC_tn}
\end{figure}

We have implemented the Jaccard-Gaussian algorithm and its performance is plotted against our method in Fig. \ref{fig:Ent_JC_tn}. The Jaccard-Gaussian algorithm relies upon a convergence condition in  order to select the width parameter used in its Gaussian weighting. We follow the convergence condition defined in the supplementary material of \cite{temporalRobust}:
\begin{equation} |P_{n}(\theta_{map}) - P_{2n}(\theta_{map})| \le 0.05,
\label{eq:JGconv}
\end{equation}
where $P_{n}$ refers to the candidate probability distribution using a stencil of $n$ simulated runs and $\theta_{map}$ refers to the most likely candidate node inside $P_{2n}$. Additionally, the two probability distributions were generated from independent simulations. We used the same set of potential Gaussian widths as \cite{temporalRobust} and selected the smallest weight for which the convergence condition is satisfied. We generated a stencil of 200 runs for different values of the transmission parameter $\lambda$ and $T = 5$ and stored these runs for testing the performance of our algorithm against the Jaccard-Gaussian algorithm. In all cases, the available information is the same for both algorithms. However, the convergence condition for the Jaccard-Gaussian method limits the stencil set from 200 to 134. Therefore, we limit our algorithm's stencils to use the same 134 simulated runs as well.

To test how the performance is affected by the stencil size, we again generated a stencil set of 1000 runs for $\lambda = 0.7$ and $T = 5$. The convergence condition limits the final stencil used to 667 runs for both methods. The spectrum of the true seed's ranking in both methods are shown in Fig. \ref{fig:Ent_JC_tn}b. From Fig. \ref{fig:Ent_JC_tn}a the probability of finding the true seed ($\lambda = 0.7$) for our method and Jaccard-Gaussian method is $0.5890 \pm 0.0419$ and $0.3653 \pm 0.0422$ respectively. When the stencil size is increased to 667 runs, these scores become $0.5776 \pm 0.0430$ and $0.3744 \pm 0.0377$ for our method and Jaccard-Gaussian method respectively. Although the Jaccard-Gaussian method should be improved with a larger stencil size, the amount of simulations required can be quite large.

We conclude our discussion with the limitations of our inversion scheme. As usual, there is a trade off between accuracy and generality. Our method should not be expected to perform better than methods that are custom tailored to specific models. By studying the specific dynamics, one can generate additional constraints and properties of the dynamics such as tree-topology, exact analytical solutions, conservation laws, etc, that might aid in inversion \cite{rumorSITreeDong, rumorSITreeShah}. Furthermore, the performance of our algorithm can be enhanced by extending what we have done with 2-body correlations to $n$-body correlations or time dependent correlations to calculate path integrals based on Bayesian inference (e.g. \cite{bayesBelief} for one specific model). However, generalizing such approaches for \emph{any} forward model and \emph{any} network topology is offset by the huge number of simulations required to resolve the correlations to within a useful error margin, and will be very costly.

%\begin{acknowledgments}
{\bf Acknowledgements.} We thank Fatos Yarman Vural for her insights. This material is based upon work supported by the Defense Advanced Research Projects Agency, HR0011-16-C-0062.
%\end{acknowledgments}

\begin{figure*}
\centering
\includegraphics[width=\linewidth]{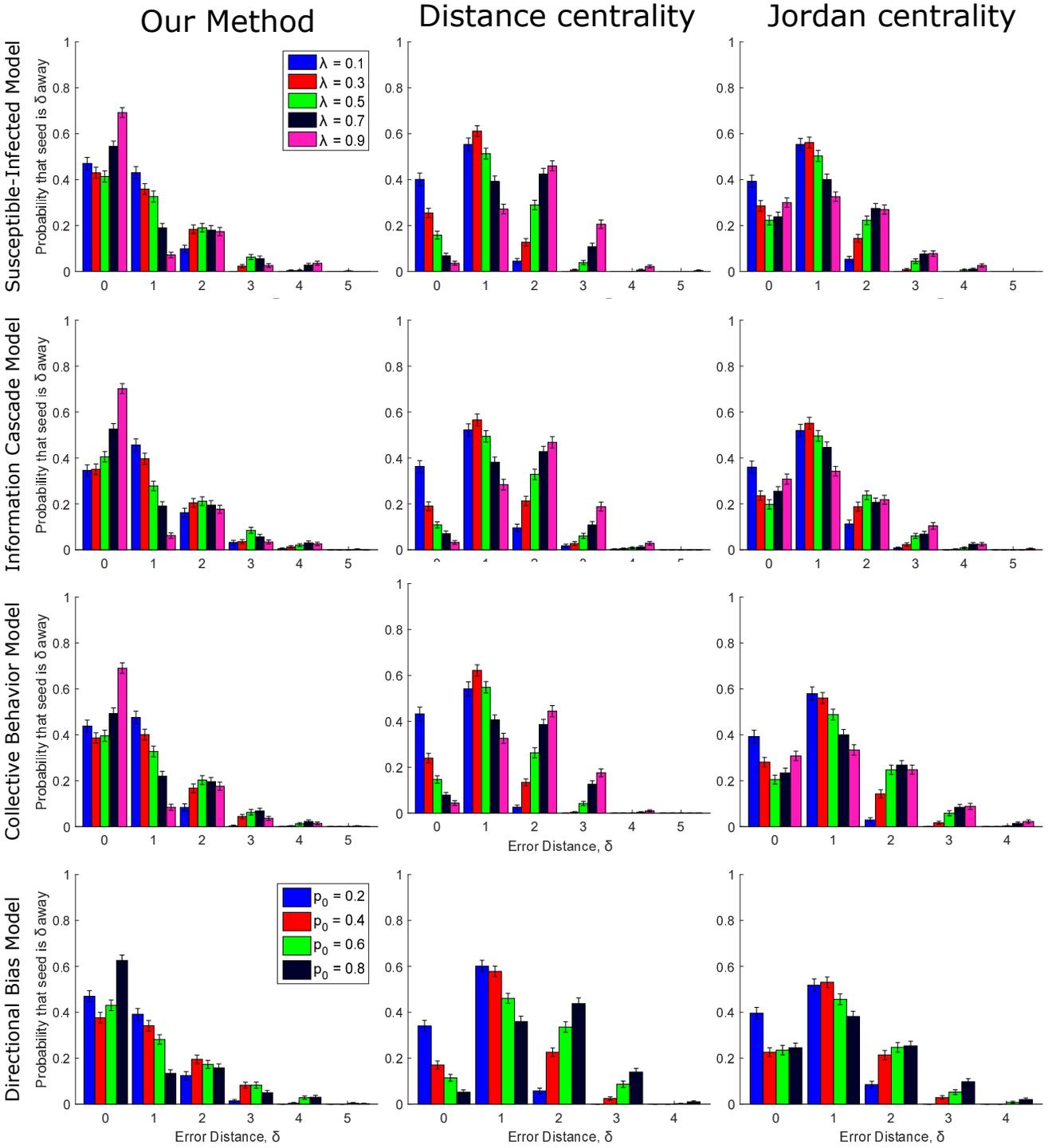}
\caption{Error distance $\delta$ for the power grid network (n = 4941, m = 6594). $\delta$ is the distance between the top candidate and the true origin.}
\label{fig:DGRID}
\end{figure*}

\begin{figure*}
\centering
\includegraphics[width=\linewidth]{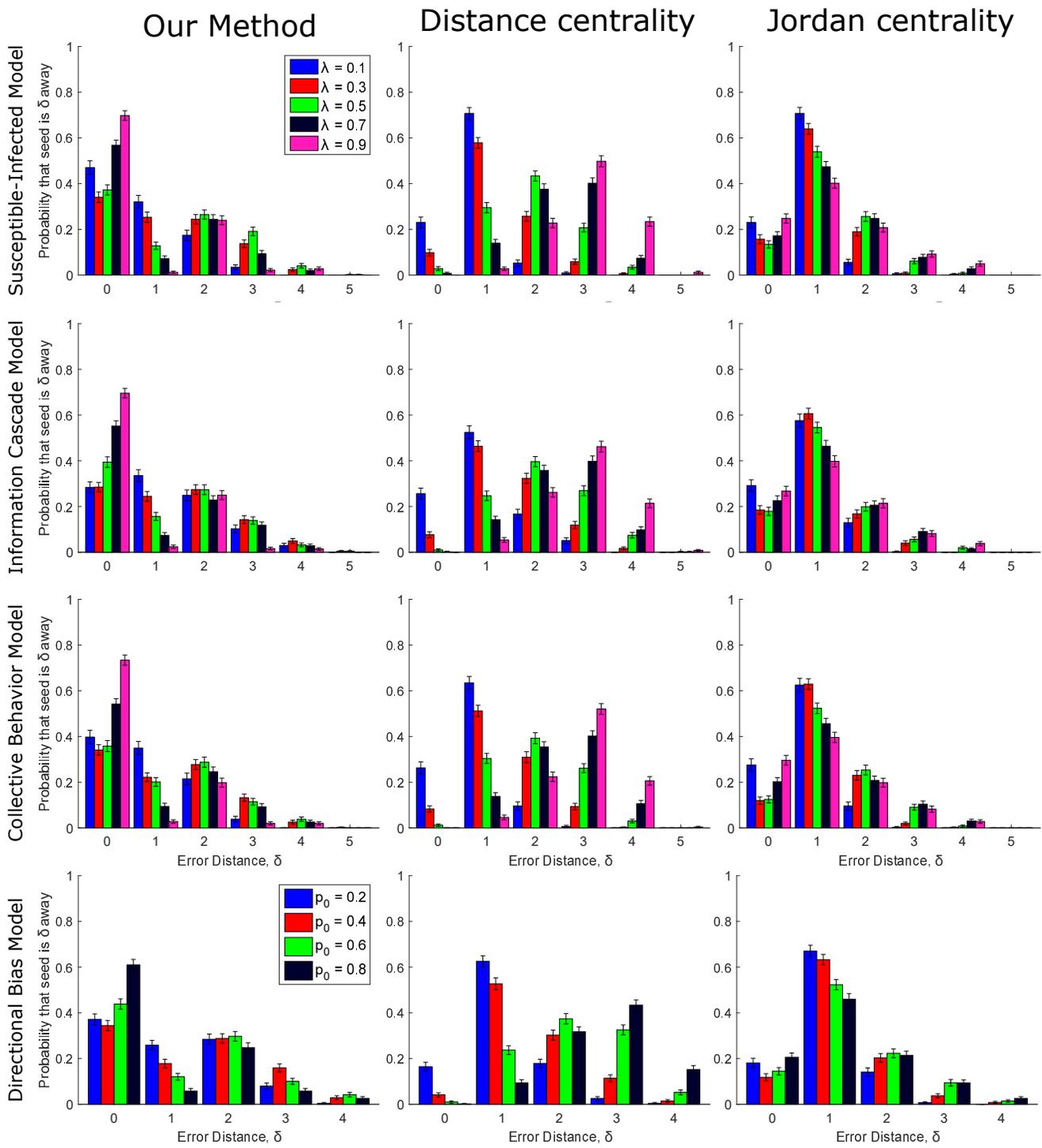}
\caption{Error distance $\delta$ for the scale-free network (n = 4941, m = 6601). $\delta$ is the distance between the top candidate and the true origin.}
\label{fig:DSCLF}
\end{figure*}

\begin{figure*}
\centering
\includegraphics[width=\linewidth]{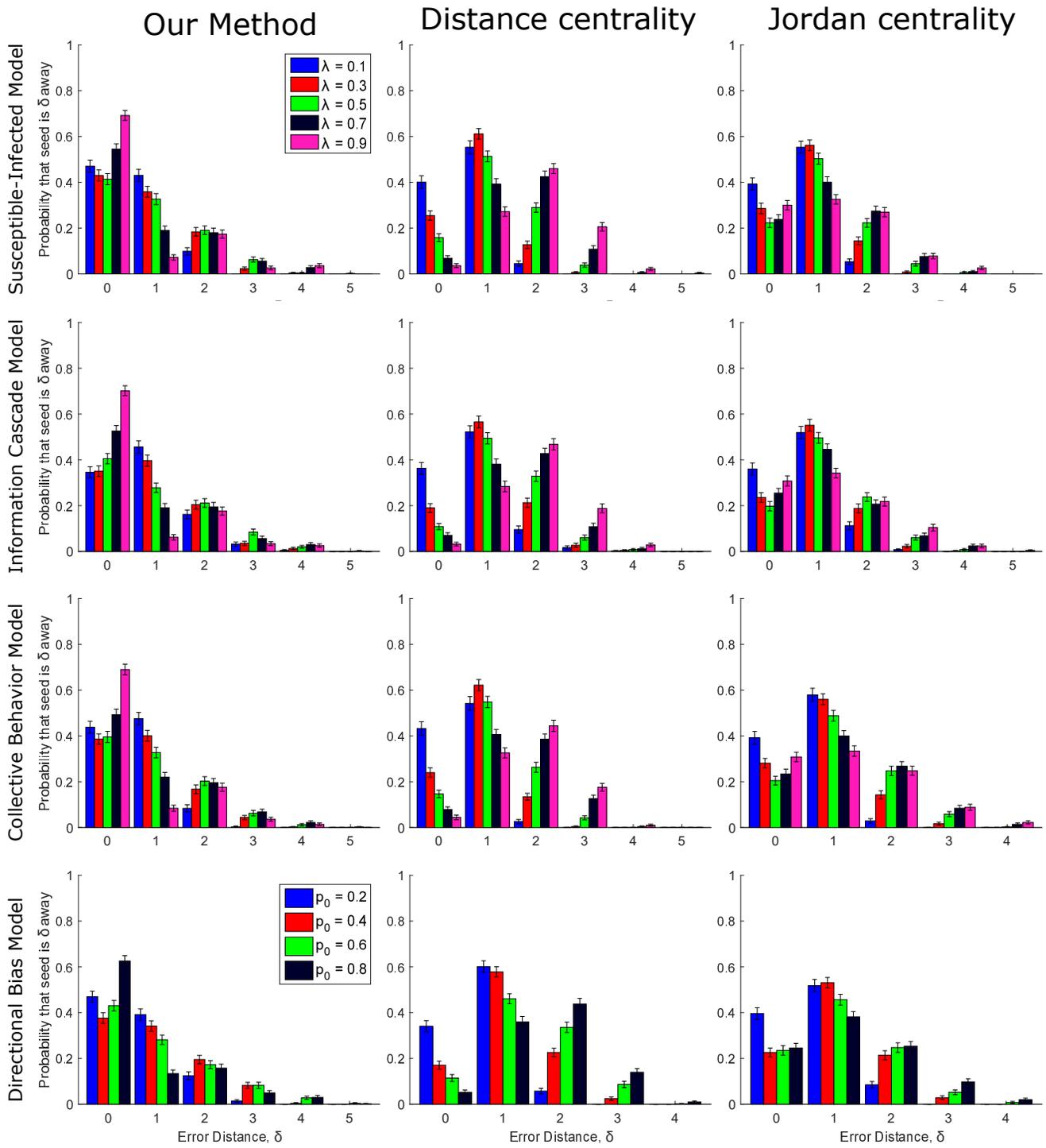}
\caption{Error distance $\delta$ for the protein interaction network (n = 3744, m = 7749). $\delta$ is the distance between the top candidate and the true origin.}
\label{fig:DPROT}
\end{figure*}

{\bf Appendix - Distance Errors.} We calculate, for the convenience of compare to other authors, the performance of our algorithm based on the distance between our top candidate and the true seed in Fig. \ref{fig:DGRID}, \ref{fig:DSCLF}, and \ref{fig:DPROT}. A distance error of zero means that we correctly identified the true seed.

\begin{figure*}
\centering
\includegraphics[width=\linewidth]{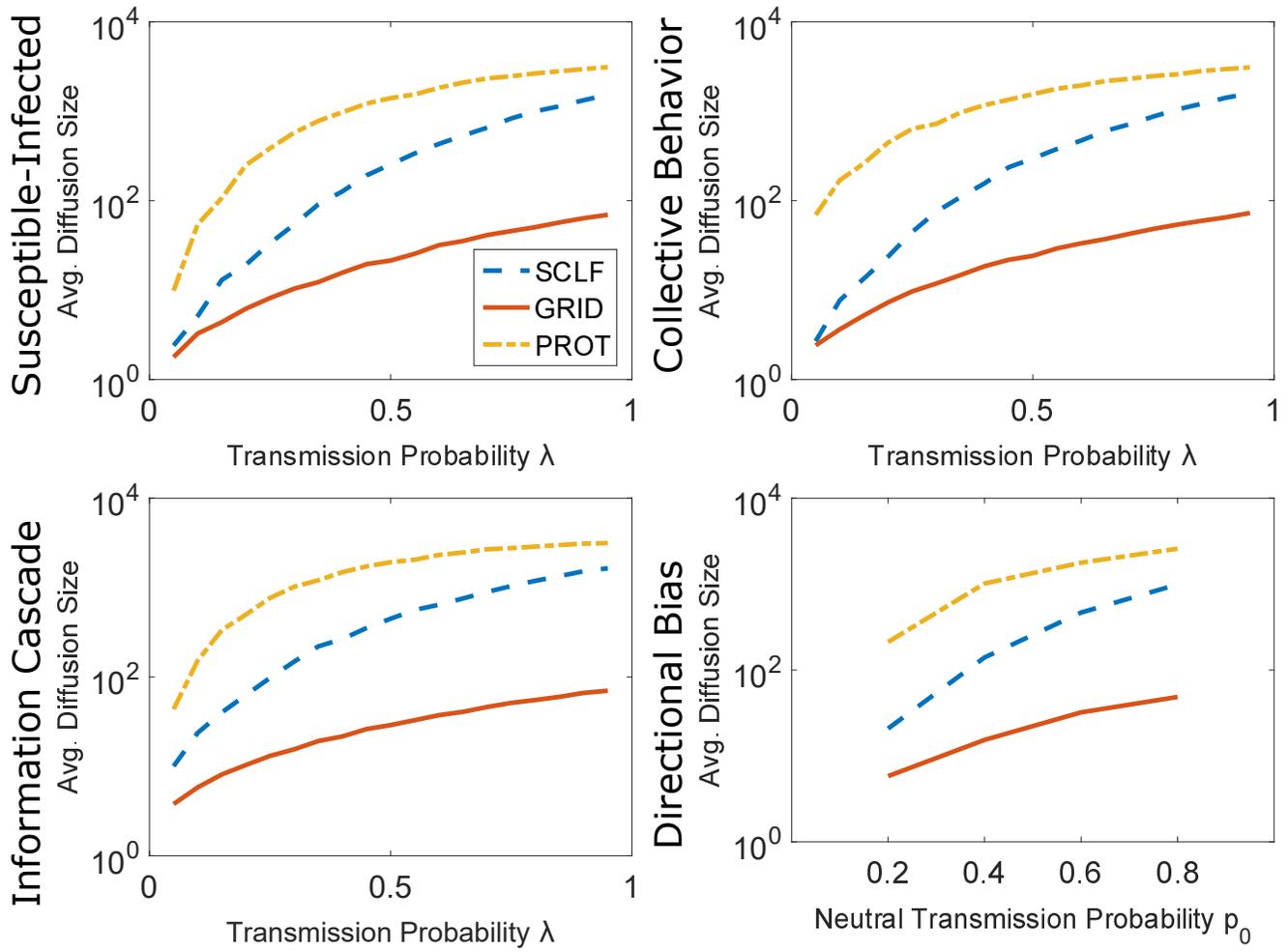}
\caption{Average Diffusion Sizes.}
\label{fig:infsize}
\end{figure*}

{\bf Appendix - Diffusion Sizes.} The average size of the diffusion is plotted for different topology, network, and transmission probabilities in Fig. \ref{fig:infsize}.

\end{document}